\documentclass{jfm}
\usepackage{graphicx}
\usepackage{natbib}

\begin{document}
\title{Bound states in the continuum in open acoustic resonators}

\abstract{We consider bound states in the continuum (BSC) or embedded trapped modes in two- and three-dimensional
acoustic axisymmetric duct-cavity structures. We demonstrate numerically that under variation of the length of the cavity
multiple BSCs occur due to the Friedrich-Wintgen two-mode full destructive interference mechanism.
The BSCs are detected by tracing the resonant widths to the points of the collapse of Fano resonances where one of
the two resonant modes acquires infinite life-time. It is shown that the approach of the acoustic coupled mode theory
cast in the truncated form of a two-mode approximation allows us to analytically predict the BSC
frequencies and shape functions to a good accuracy in both two and three dimensions.}

\author[A. A. Lyapina, D. N. Maksimov, A. S. Pilipchuk, and A. F. Sadreev,]
{A. A. Lyapina$^{1,2}$,\ns D. N. Maksimov$^1$\thanks{Email address for correspondence: mdn@tnp.krasn.ru},\ns, A. S. Pilipchuk$^{1,2}$ and
A. F. Sadreev$^1$,\ns \break }
\affiliation{$^1$L.V. Kirensky Institute of Physics, Krasnoyarsk, 660036, Russia \\[\affilskip]
$^2$Siberian Federal University, Krasnoyarsk, 660041, Russia}
\maketitle

\section{Introduction}

Bound states in the continuum (BSCs) also known as embedded trapped modes are localized solutions
which correspond to discrete eigenvalues coexisting
with extended modes of a continuous spectrum in resonator-waveguide configurations. The existence of
trapped solutions residing in the continuum
was first reported by \cite{Neumann1929} at the dawn of quantum mechanics. To the best of our knowledge,
the term {\it bound state (embedded) in the continuum} was introduced by \cite{Fonda1961} in the context of
resonance reactions in the presence continuous channels. Since then {\it bound state in the continuum} has been
universally used to designate a BSC in quantum mechanics \cite{Stillinger1975}. In the field of fluid mechanics,
\cite{Parker66, Parker1967} is credited to be the first to encounter resonances of pure acoustic
nature in air flow over a cascade of flat parallel plates.
Nowadays, the BSCs are known to exist in various waveguide structures
ranging from quantum wires \cite{Shahbazyan1994, Kim1999, Olendski, Sadreev2006, Cattapan},
acoustic waveguides \cite{Linton2007,Duan2007, Hein2008, Hein2012}, and photonic
crystals \cite{Shipman, Bulgakov2008, Marinica2008, Yang}. The BSCs are of immense interest,
specifically, in optics thanks to experimental opportunity to confine light in optical microcavities
despite that outgoing waves are allowed in the surrounding medium
\cite{Plotnik2011,Longhi,Kivshar,WeiHsu2013,Zhang}. At the same time, in aerodynamics, trapped and nearly trapped modes
are known to cause severe vibrations
and noise problems in gas and steam pipelines \cite{Jungowski1989, Ziada1992, Kriesels1995,
Ziada1999, Tonon2011}.

The most trivial mechanism of BSC formation is due to the symmetry of the structure i.e. the BSC and
the continuous-spectrum modes have incompatible symmetries. Typically, when
symmetric and antisymmetric problems are separated the first symmetric and antisymmetric propagating modes have
different cut-off frequencies. Thus, the difference between the cut-off frequencies provides a window for
an antisymmetric BSC to reside in. Such BSCs are omnipresent in wave-related set-ups including quantum wires \cite{Schult1989}, fluid
in a wave tank \cite{Evans1991}, acoustic \cite{Evans1994, Sugimoto2005} and elastic
\cite{Maksimov2006} waveguides.


By breaking the axial symmetry of the structure the symmetry protected BSCs become quasi-trapped leaky modes
coupled to the propagating modes of the waveguides \cite{Aslanyan2000}. \cite{Nockel92, Guevara2003} demonstrated that if the
symmetry is broken under variation of a control parameter the symmetry protected BSCs reveal themselves
in form of ghost Fano resonances. Conversely, as the symmetry is recovered
the Fano resonance collapses with one of the two resonant modes acquiring an infinite live-time.
The collapse of Fano resonance is a signature of full destructive interference of two degenerate eigenmodes of the
resonator on the waveguide-resonator interface \cite{Kim1999}. In the consequence of this a trapped mode is
formed as a localized solution decoupled from the extended modes of
the waveguides. Importantly, the full destructive interference of two degenerate eigenmodes leaking into waveguides represents
a generic mechanism of BSC formation \cite{Friedrich1985} whose implementations go far beyond
the above symmetry arguments by \cite{Guevara2003}.
In particular, this mechanism allows the formation of BSCs with the same
symmetry about the duct axis as the coexisting scattering solution \cite{Hein2012}. The BSCs which can be attributed to
a full destructive interference
are multiply reported in quantum mechanics \cite{Kim1999},
optics \cite{Bulgakov2014}, and acoustics \cite{Hein2010, Hein2012}. The same interference mechanism was recently employed
by \cite{Lepetit1, Lepetit2014} for
experimental observation of BSCs in microwave set-ups. Throughout this paper the full destructive interference of two degenerate modes
resulting in the formation of BSCs in multimodal cavities will be invoked to establish {\it two-mode approximation} that
will be used for finding BSC frequencies and shape functions.


In our recent work \cite{Maksimov2015}
we showed that the approach known in quantum mechanics as formalism of the effective non-Hermitian Hamiltonian \cite{Dittes2000,Pichugin}
could be adapted for solving hard-wall acoustic cavity-duct problem. The essential feature of the approach is that
it allows to recover transmission spectra of open-boundary systems relying on the spectral
properties of their closed counterparts. Technically, the method is the acoustic coupled mode theory in
which the pressure field within the cavity is represented by the modal variables corresponding to the amplitudes of
the eigenmodes of the closed cavity decoupled from waveguides. We also address the reader to the paper by \cite{Racec09}
for a similar approach for scattering in cylindrical nanowire heterostructures.


In this paper, we consider acoustical problem of axisymmetric side-branch cavities in two- and three-dimensional
ducts of infinite length. Our primary goal is to explore the interference phenomena
resulting in the formation of BSCs localized within two- and three-dimensional cavities.
The BSCs will be analysed in the framework of the acoustic coupled mode theory \cite{Maksimov2015}.
We will demonstrate that our approach not only represents an alternative numerical technique for finding BSCs but also
provides an efficient
analytical tool for construction the BSC shape function in the two-mode approximation. In particular, we will
see that the two-mode approximation is able to a good accuracy reproduce the BSC pressure field within two- and three-dimensional
acoustic cavities.


The paper is organized as follows. In Section 2 we briefly overview the acoustic coupled mode theory which is our major
tool for analysing the scattering problem. The numerical techniques for finding BSCs are also discussed. In Section 3
we present our results on BSCs in two-dimensional duct-cavity systems. Three-dimensional systems are considered in Section 4.
Finally, we conclude in Section 5.

\section{Acoustic Coupled Mode Theory}

Let us start with the formulation of the problem. Throughout this paper we adopt non-dimensionalized quantities.
After the time dependance is removed
through the substitution $\psi(t)=\psi \exp(-i\omega t)$ the pressure field $\psi$ within a cavity-duct structure is
controlled by the non-dimensional stationary Helmholtz equation
\begin{equation}\label{Helm}
\left(\frac{\partial^2}{\partial x^2} + \frac{\partial^2}{\partial y^2} + \frac{\partial^2}{\partial z^2}  \right)\psi+ \omega^2\psi=0,
\end{equation}
where $\omega$ is the non-dimensional frequency. The $x$-axis is aligned with
the center line of the duct-cavity structure. The problem could be reduced to two dimensions under assumption
that the structure has constant thickness along the $z$-axis which
allows us to present the solution of Eq. (\ref{Helm}) as $\psi=\psi(x,y)\psi(z)$. In what follows we take $\psi(z)=const$ which means
that only the first propagating mode is involved.

Eq. (\ref{Helm}) has to be solved subject to Neumann boundary conditions on all sound-hard boundaries of the cavity and reflectionless
boundary conditions on the cavity-waveguide interfaces. There are numerous numerical techniques to impose reflectionless conditions on
waveguide-cavity interfaces the most efficient of which implement layers of absorbing  medium with complex refractivity giving rise to
{\it perfectly matched layers} \cite{Berenger1994} or {\it complex scaling} \cite{Moiseyev1998} methods. For detailed information on the use
of those methods for finding
acoustic BSCs we address the reader to the papers by \cite{Hein2008, Hein2012} which thoroughly review the subject. In this
paper we will use an alternative method, namely, {\it acoustic coupled mode theory} \cite{Maksimov2015}. The method is a recent adaptation of
the approach known in quantum mechanics as {\it formalism of the effective non-Hermitian Hamiltonian}. In this section we only briefly
review the essentials and introduce the basic concepts that will be needed later on for understanding the mechanism of BSC formation.

\begin{figure}
\begin{center}
\includegraphics[width=0.8\textwidth]{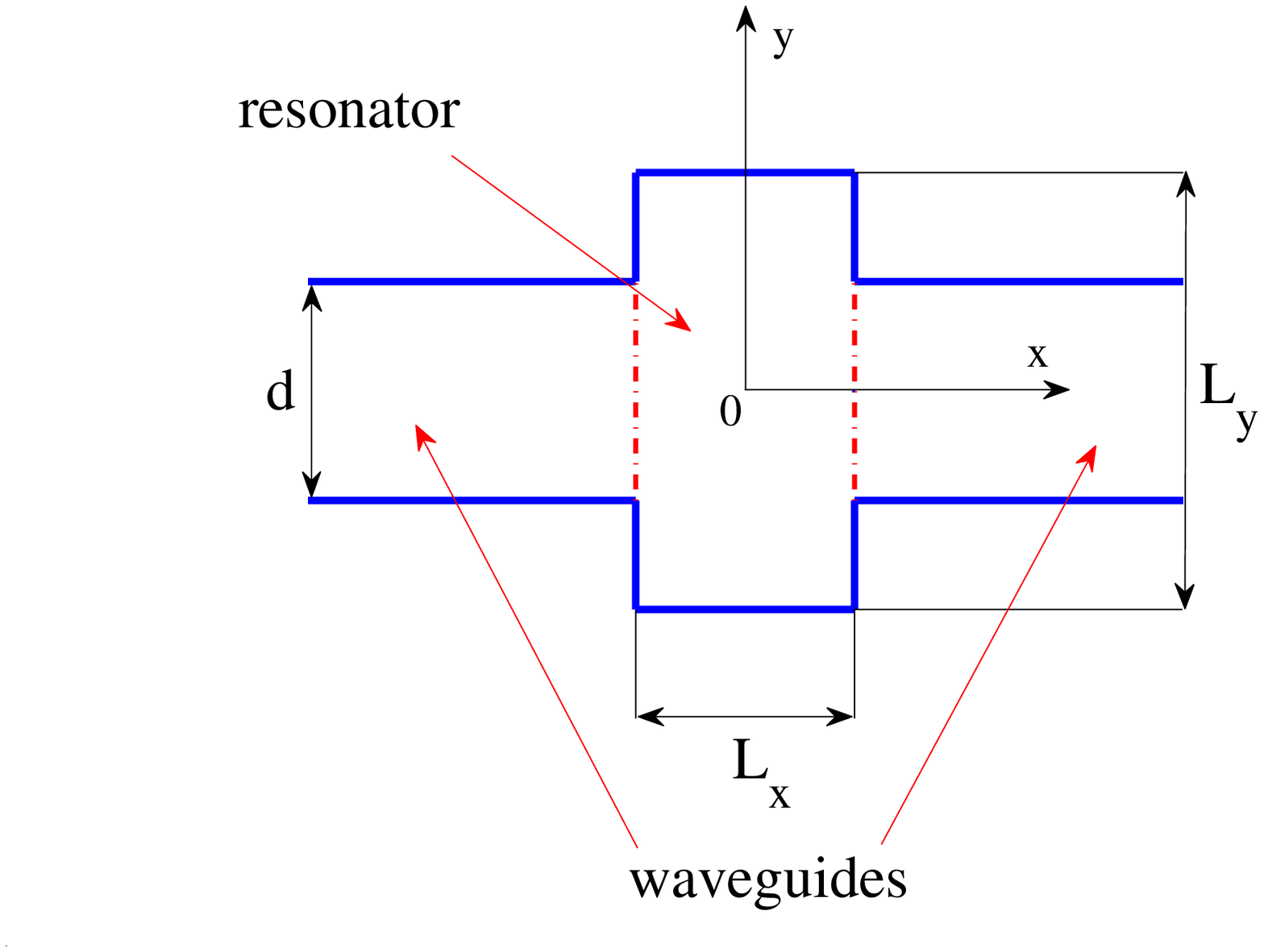}
\caption{(Color on-line) Structure lay-out. $d=1$.}
\label{fig:lay-out}
\end{center}
\end{figure}

To be consistent with the further applications  our approach is exemplified on the two-dimensional duct-cavity structure depicted in
Fig. {\ref{fig:lay-out}} where we show a two-dimensional cavity of length $L_x$ and width $L_y$ coupled to symmetrically
positioned
waveguides of width $d$. In what follows we take $d=1$.
As it was mentioned in the Introduction the coupled mode theory relies
on the spectral properties of
a closed cavity decoupled from waveguides. In that context {\it closed} means that in the first step the eigenmodes and eigenfrequencies
of the cavity are computed with the Neumann boundary conditions
$${\left. \frac{\partial\psi}{\partial \sigma}\right|}_{\partial\Omega}=0 $$
on the boundary of the resonator $\partial\Omega$ including the waveguide-cavity interfaces as well as the
physical boundaries; $\sigma$ being the local coordinate normal the boundary of the closed resonator. In the case of a rectangular
cavity the eigenvalue problem is easily solved analytically to yield the spectrum as
\begin{equation}\label{2D_spectrum}
\omega_{n,m}^2={\left(\frac{\pi (n-1)}{L_x}\right)}^2+{\left(\frac{\pi (m-1)}{L_y}\right)}^2; \ n,m=1,2,\ldots
\end{equation}
At the same time the eigenmodes are found as products of two factors
\begin{equation}\label{2D_eigenmodes}
\psi_{n,m}=\sqrt{\frac{(2-\delta^1_n)(2-\delta^1_m)}{L_xL_y}}
\cos{\left(\frac{\pi (n-1)(2x+L_x)}{2L_x}\right)} \cos{\left(\frac{\pi (m-1)(2y+L_y)}{2L_y}\right)},
\end{equation}
where $\delta^{n}_{n'}$ is the Kronecker delta.
Notice that the eigenfunctions are unit-normalized, so that
\begin{equation}
\int\psi_{n,m}^2dxdy=1.
\end{equation}

The propagating solutions in the waveguides have a form of plane waves
\begin{equation}\label{2D_waveguides}
\phi_p^{(\pm)}(x,y)=\sqrt{\frac{1}{4\pi k_p}}\exp{(\pm ik_px)}\chi_p(y),
\end{equation}
where symbol $+(-)$ stands for the waves propagating to the right (left), and the functions $\chi_p(y)$
are defined as
\begin{equation}\label{chi}
\chi_p(y)=\sqrt{2-\delta^1_p}\cos\left(\pi \frac{(p-1)(2y+1)}{2}\right).
\end{equation}
The dispersion relation reads
\begin{equation}\label{dispersion}
k_p=\sqrt{ \omega^2-{\pi^2\left( p-1\right)}^2}; \ p=1,2,\ldots.
\end{equation}

The pressure field within the cavity $\psi_{c}$ could be constructed as a modal expansion
\begin{equation}\label{expansion}
\psi_{c}=\sum_{n,m}g_{n,m}\psi_{n,m}.
\end{equation}
over eigenfunctions Eq. (\ref{2D_eigenmodes}).
The key equation for finding the unknowns $g_{n,m}$ for the cavity excited by an incoming wave has the following form
{\cite{Maksimov2015}}
\begin{equation}\label{CMT}
(D-\omega^2I){\bf g}=-i\sum_{p=1}^{\infty}\sqrt{\frac{k_p}{\pi}}\sum_{C=L,R}{\bf v}_{C,p} a_{C,p}
\end{equation}
where $I$ is the identity matrix, the subscript $C=L,R$ specifies the coupling with the
left and right waveguides, and the unknown vector $\bf g$ contains the coefficients $g_{n,m}$.
Analogously, the amplitudes ${a}_{C,p}$ are the coefficients of the modal expansion of the incident wave field
in the waveguides over the eigenfunctions (\ref{2D_waveguides}), while the vector ${\bf v}_{C,p}$ contains the coupling constants
$v_{C,p}^{m,n}$ between the specific eigenmode $(m,n)$  (\ref{2D_eigenmodes}) and $(C,p)$ propagating mode of the waveguides
(\ref{2D_waveguides}). Finally, the matrix $D$ has the following form
\begin{equation}\label{Heff}
D=\Omega^2-\sum_p^{\infty}ik_p\sum_{C=L,R}{\bf v}^{\dagger}_{C,p}{\bf v}_{C,p},
\end{equation}
where $\Omega^2$ is a diagonal matrix which carries squared eigenfrequencies $\omega_{n,m}^2$ Eq. (\ref{2D_spectrum}) on the
main diagonal and the symbol $\dagger$ stands for Hermitian transpose.
Notice that the coupling to the evanescent modes $(p-1)\pi>\omega$ of the waveguides
is accounted for in
Eq. (\ref{Heff}) by the terms with imaginary wavenumbers $k_p$ found from Eq. (\ref{dispersion}).
At this point we only have to specify the coupling constants
$v_{C,p}^{n,m}$. According to \cite{Pichugin, Maksimov2015} they are evaluated as overlapping integrals between the functions
(\ref{chi}) and the eigenfunctions (\ref{2D_eigenmodes}) on the waveguide-cavity interface
\begin{equation}\label{coupling}
v_{C,p}^{m,n}=\int \chi_{p}\psi_{m,n}d\tau_{C},
\end{equation}
where $\tau_{C}$ is the local interface coordinate orthogonal to the wave vector. The generalization to the three-dimensional case
is straightforward; the coupling constant is a surface integral \cite{Maksimov2015} with $\tau_C$ representing the local coordinate
 set
on a flat waveguide-resonator interface.
Notice that Eq. (\ref{CMT}) formally contains an infinite
number of unknowns. However, as it was demonstrated in \cite{Maksimov2015} for computational sake the basis can be truncated to
a reasonably small number of modes in a given
frequency window. After the unknown vector $\bf g$ is obtained from Eq. (\ref{CMT}) the coefficients of
the modal expansion of the outgoing
field in the waveguide $b_{C,p}$ are found as
\begin{equation}
b_{C,p}=-a_{C,p}+\sqrt{4\pi k_p} {\bf v}^{\dagger}_{C,p}{\bf g}.
\end{equation}

The above described approach is demonstrated to be applicable for finding scattering functions and reflection amplitudes in
two- and three-dimensional cavities \cite{Maksimov2015}. The goal of the present paper, however, is to adapt it for computing BSCs. The BSCs
are localized solutions
of infinite live-time which exist in the cavity even when no incident field is present in the waveguides. Thus, according to
Eq. (\ref{CMT})
one obtains an eigenvalue problem
\begin{equation}\label{BSCs}
(D-z^2_nI){\bf g_n}=0.
\end{equation}
Once, an eigenvalue $z_n=\omega_n+i\gamma_n$ found from Eq. (\ref{BSCs}) is real its eigenvector $\bf g_n$ corresponds to an BSC.
Unfortunately, Eq. (\ref{BSCs}) does not represent the standard eigenvalue problem because the
matrix $D$ Eq. (\ref{Heff}) depends on the spectral parameters $k_p$ which are linked to the
eigenvalue $z_n$ through the dispersion relation Eq. (\ref{dispersion}).
One of the possible solutions \cite{Sadreev2006} is employing an iterative procedure of solving the fix-point equation
for real
eigenvalues of Eq. (\ref{BSCs}). Alternatively, one of the most efficient techniques for finding complex eigenvalues
(\ref{BSCs}) is the
{\it harmonic inversion method} which allows to extract the positions and life-times of the resonances from the response of the system to
an external driving. In this paper we do not detail the Harmonic Inversion owing the credit to \cite{Wiersig2008} who nicely outlined
the method in their paper on the fractal Weyl law for chaotic microcavities. In brief, the first step in the method is solving Eq. (\ref{CMT})
with some, quite arbitrary, right hand part which, in particular, could be a point source placed within the open cavity. The response obtained
is then Fourier-transformed to yield a set of nonlinear equation for eigenvalues $z_n$ which is subsequently solved by the use of Pad{\'e}
approximants
\cite{Wiersig2008}.

To conclude this section we return to the Friedrich-Wintgen two-mode interference mechanism of BSC
formation \cite{Friedrich1985}. Let us check whether it is possible that only
two modes $\psi_{1}$ and $\psi_2$ of the closed
cavity with eigenfrequencies $\omega_1$ and $\omega_2$ contribute to a BSC. Suppose that only the first scattering channel with the wave
number $k_1$ is open at the frequency of interest. Then, under a further assumptions that the coupling to the evanescent channels
$p>1$ could be neglected we rewrite Eq. (\ref{BSCs}) as an
eigenvalue problem for matrix $D$ with real valued terms $\omega_j^2$ arranged along the main diagonal in the ascending order
\begin{equation}\label{two_mode}
D=\left(\begin{array}{cccccc}
\ddots & \vdots & \vdots & \vdots & \vdots & \ddots\cr
\ldots & \omega_0^2-ik_12v_0^2 & -ik_12v_0v_1 & -ik_12v_0v_2 & -ik_12v_0v_3 & \ldots \cr
\ldots & -ik_12v_0v_1 & \omega_1^2-ik_12v_1^2 & -ik_12v_1v_2 & -ik_12v_1v_3 & \ldots \cr
\ldots & -ik_12v_0v_2 & -ik_12v_1v_2 &  \omega_2^2-ik_12v_2^2 & -ik_12v_2v_3 & \ldots \cr
\ldots & -ik_12v_0v_3 & -ik_12v_1v_3 &  -ik_12v_2v_3 & \omega_3^2-ik_12v_3^2 & \ldots \cr
\ddots & \vdots & \vdots & \vdots & \vdots & \ddots
\end{array}\right),
\end{equation}
where $\omega_0$ and $\omega_3$ are the eigenfrequencies next to $\omega_1$ and $\omega_2$, and $v_j=v^{m,n}_{L,1}$ is the coupling
constant with the open propagation channel $p=1$.
Notice, that above we assumed that the absolute values of the coupling constants on the left and the right are identical
i.e. all eigenmodes have compatible symmetry about the $y$-axis. In fact, the matrix $D$ is split into two decoupled subblocks
corresponding to the $y$-symmetric and $y$-antisymmetric eigenmodes so that the $y$-symmetric and $y$-antisymmetric problems
could be considered independently from one another.
Under variation of some control parameter, say the length of the cavity $L_x$, one
can reach the degeneracy point where $\omega_1=\omega_2=\omega_{BSC}$, then the matrix (\ref{two_mode}) has a real eigenvalue
 $\omega^2_{BSC}$
with the eigenvector
$${\bf g}= \frac{v_1v_2}{\sqrt{v_1^2+v_2^2}}{\left(\begin{array}{cccccc} \ldots, & 0, & 1/v_1, & -1/v_2, & 0, &  \ldots \end{array}\right)}^{\dagger}$$
containing only two non-zero entries. Thus, the corresponding BSC shape-function can be constructed as
\begin{equation}\label{Shape_function}
\psi_{BSC}=g(1)\psi_1+g(2)\psi_2.
\end{equation}
The above equation represents the central mathematical result to be used for analyzing BSCs in open acoustic resonators. In essence, it manifests
the well known interference mechanism \cite{Friedrich1985}. In fact, both eigenmodes $\psi_{1}$ and $\psi_2$ albeit coupled
to the waveguides contribute to Eq. (\ref{Shape_function}) in the proportion providing a full destructive interference at the
degeneracy point so that the resulting BSC function $\psi_{BSC}$ is decoupled from the open channel. For details on the realization
of the above mechanism in the field of quantum mechanics we address the reader to the paper by \cite{Sadreev2006}. We mention in passing
that thanks to the symmetry about the duct axis the same arguments are valid for the duct-axis antisymmetric modes with $k_1$ replaced by $k_2$
according to Eq. (\ref{dispersion}).

\begin{table}
\begin{center}
\begin{tabular}{lcccccccc}
p \textbackslash (n,m) & (1,1) & (1,2) & (1,3) & (1,4) & (4,1) & (4,2) & (2,3) & (2,4) \\
\hline
1 & 0.408 & 0     & -0.367 & 0      & 0.577 & 0     & -0.520 &  0       \\
2 & 0     & 0.245 & 0      & -0.441 & 0     & 0.346 & 0      &  -0.624  \\
3 & 0     & 0     & 0.173  & 0      & 0     & 0     & 0.245  &  0       \\
4 & 0     & 0.021 & 0      & 0.082  & 0     & 0.030 & 0      &  0.116   \\
5 & 0     & 0     & 0.034  & 0      & 0     & 0     & 0.049  &  0       \\
6 & 0     & 0.007 & 0      & 0.024  & 0     & 0.011 & 0      &  0.034   \\
\end{tabular}
\caption{Numerical values of coupling constants $v_{L,p}^{n,m}$ at $L_x=3, L_y=2$.}
\label{Table}
\end{center}
\end{table}

The above result Eq. (\ref{Shape_function}) is exact under assumption that the coupling to the evanescent channels could be neglected.
However, we would like to remind the reader that the evanescent eigenmodes are present in summation over index $p$ in
the right hand part of Eq. (\ref{Heff}). Therefore Eq. (\ref{Shape_function}) is not an exact solution, and could only be seen as the basis for a
{\it two-mode approximation}
for BSCs occurring at the degeneracy points. Some remarks are due on the validity of the two-mode approximation.
As it will be seen later on due to the structure of Eq. (\ref{Heff}) the evanescent
coupling results in the presence of other than the two basic modes in the modal expansion of the BSC shape function, as well as in
a shift of the true BSC point from the degeneracy point. Nevertheless, one can argue that the coupling to the evanescent channels
is small relative to
the coupling to the open channel as
we consider the eigenmodes residing in the frequency window where only one symmetric and one antisymmetric propagating
channels are open. In order to illustrate this we collected in Table \ref{Table} the numerical values of the coupling constants for some relevant
eigenmodes evaluated on the basis of the analytical results from \cite{Maksimov2015}. One can see from Table \ref{Table} that the
coupling constants rapidly decay with the growth
of the channel number $p$. In fact, in can be demonstrated that any coupling constant decays away from its maximal value with the asymptotic
law $v_{L,p}^{n,m}\sim 1/p^2$.

Finally, notice that the two-mode approximation has no
restrictions related to the size of the cavity. One can see from Eq. (\ref{two_mode}) that once a degeneracy occurs in the closed cavity
it always results in the formation of a BSC through the Friedrich-Wintgen mechanism irrelevant to the values of $L_x$, and $L_y$.
If the size of the cavity is isometrically increased whilst the width of the waveguide is kept unchanged, the next to the BSC
eigenfrequencies $\omega_0, \omega_3$ would shift closer to the BSC eigenfrequency. The real valued diagonal term in Eq. (\ref{BSCs}) will drop as
$(\omega_{0,3}^2-\omega_{BSC}^2)\sim 1/S, S=L_xL_y$ because the mean distance between the eigenfrequencies is inversely proportional
to the width/length of the cavity. On the other hand the coupling constants $v_j$ in matrix $D$ (\ref{two_mode}) will drop as
$v_j\sim 1/\sqrt{S}$ because the eigenfunctions Eq.(\ref{2D_eigenmodes}) contributing to the overlapping integrals Eq. (\ref{coupling}) have
the normalization constant $\sqrt{S}$ in denominator. In consequence of this the off-diagonal terms $v_jv_{j'}$ in matrix (\ref{two_mode})
drop as $v_jv_{j'}\sim 1/S$. Therefore, although in larger cavities the next to the BSC eigenvalues appear closer to
the BSC point their coupling
to the basic BSC modes through the evanescent channels drops at the same rate as the resonant term $\omega_{0,3}^2-\omega_{BSC}^2$. This leads us to a conclusion that the two-mode approximation
does non break down as the size of the cavity is increased. Notice, that the same holds true in the three-dimensional case
where the coupling constant drops as $v_j\sim 1/\sqrt{V}$ with $V$ as the volume of the cavity.

\begin{figure}
\begin{center}
\includegraphics[width=1\textwidth]{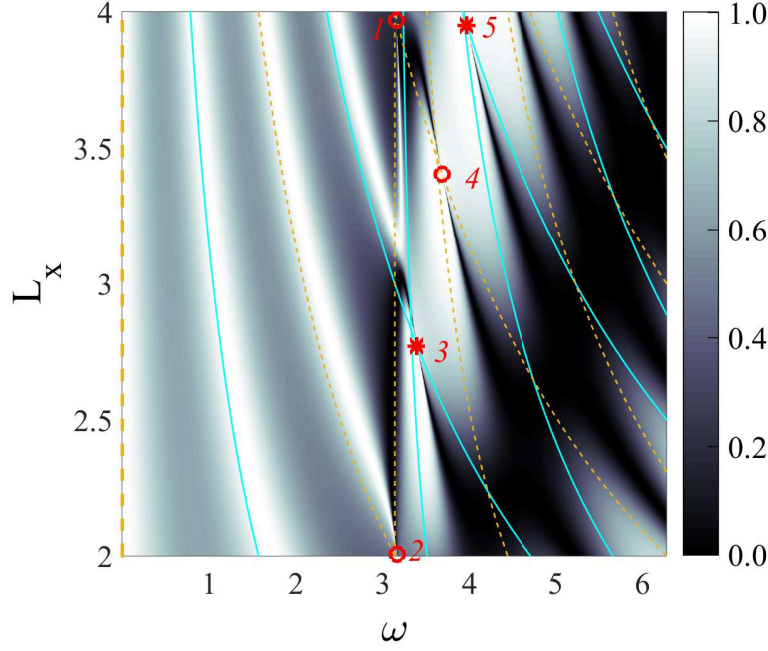}
\caption{(Color on-line) Transmittance for a two-dimensional cavity of length $L_y=2$. Only the first symmetric channel with $p=1$
is open. The dashed lines show eigenfrequencies of the modes symmetric about
the $y$-axis; solid line - antisymmetric ones. The positions of the BSCs are shown by open
circles and stars for $y$-symmetric and $y$-antisymmetric
modes, correspondingly.}
\label{Trans}
\end{center}
\end{figure}

\section{Bound states in two dimensional duct-cavity structure}

\begin{figure}
\begin{center}
\includegraphics[width=0.8\textwidth]{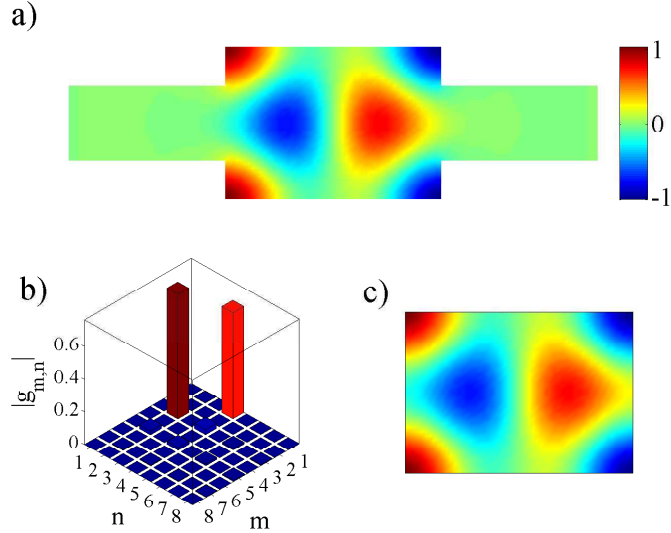}
\caption{(Color on-line) BSC {\it 3} symmetric about the duct axis of a two-dimensional duct-cavity structure
$\omega_{BSC}=3.396, L_y=2, L_x=2.776$;
a) Pressure field of the
numerically computed BSC shape function; b) Absolute values of the modal coefficient of the numerically computed pressure field;
c) BSC shape function found from the two-mode approximation Eq. (\ref{Shape_function}).}
\label{BSC1}
\end{center}
\end{figure}

We start with computing the transmittance $T={|b_{1,R}|}^2$ for the two-dimensional cavity in Fig. \ref{fig:lay-out} under assumption
that an incident wave in the first propagation band $p=1$ enters the cavity from the left waveguide.
For solving the scattering problem we used the approach described in Section 2.
The numerical values of the coupling constants $v_{C,p}^{n,m}$
in Eq. (\ref{CMT},\ref{Heff}) are taken from our previous work \cite{Maksimov2015}.
In Fig. \ref{Trans} we demonstrate the dependence of the transmittance $T$ on the frequency of the incoming wave $\omega$ and the length of the cavity $L_x$.
Notice that the frequency varies in the range $[0, {2\pi}]$ which means according to Eq. ({\ref{dispersion}}) that only one symmetric
scattering channel is open in each waveguide. In Fig. \ref{Trans} we also depict the frequencies of the eigenmodes of the closed cavity
symmetric about the duct axis Eq. (\ref{2D_spectrum}) $m=1, 3, 5, \ldots$. The eigenfrequencies which
correspond to the modes symmetric about the $y$-axis, later to be
referred to as $y$-symmetric, are shown
by dashed lines.
\begin{figure}
\begin{center}
\includegraphics[width=0.8\textwidth]{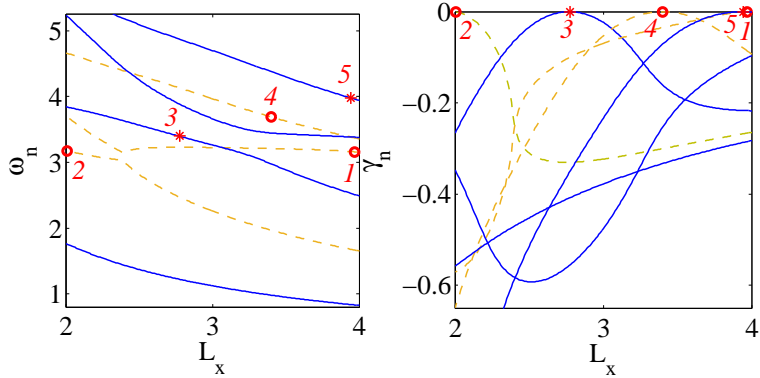}
\caption{(Color on-line) Complex eigenfrequencies against the length of the two-dimensional resonator $L_x$; real parts -
left panel;
imaginary parts - right panel. The $y$-symmetric resonances are shown by dash brown lines; $y$-antisymmetric by blue
solid lines.  The positions of BSCs from Fig. \ref{Trans} are shown by red open circles and stars for $y$-symmetric and
$y$-antisymmetric modes, correspondingly.}
\label{complex_eigenfrequencies}
\end{center}
\end{figure}
The solid lines show the eigenfrequencies of $y$-antisymmetric eigenmodes. One can see that the degeneracy
points of eigenfrequencies with the same symmetry about  the $y$-axis adjoin to an abrupt change of the transmittance;
 the transmittance drops from unit to zero as the frequency is swept across the point where the collapse of Fano resonance occurs. The complex resonance frequencies
for the cavity are computed by harmonic inversion method.
In the given range of parameters we found five BSCs. Three of them originate from
the modes symmetric about the $y$-axis (shown by circles) and two from the antisymmetric ones (shown by stars).
For further convenience we numerate the BSCs from $\it 1$ to $\it 5$. The complex eigenfrequencies
are plotted against the length of the resonator $L_x$ in Fig. \ref{complex_eigenfrequencies}  where one can see that the imaginary part
$\gamma_n$ turns to zero as one approaches the BSC point.  In Fig. {\ref{BSC1}} we demonstrate the numerical solution for
BSC $\it 3$ along with the corresponding coefficients of the modal expansion (\ref{expansion}).
One can clearly see that only two eigenmodes dominate in the BSC function $\psi_{BSC}$.
The $y$-antisymmetric BSC {\it 3} in Fig. \ref{BSC1} is formed by eigenmodes $\psi_1=\psi_{4,1}(x,y)$, and
$\psi_2=\psi_{2,3}(x,y)$ with the corresponding coupling constants $v_1=v^{4,1}_{L,1}=-v^{4,1}_{R,1}=0.600$, and
$v_2=v^{2,3}_{L,1}=-v^{2,3}_{R,1}=-0.540$. To assess the accuracy of the two-mode approximation
we introduce parameters
\begin{equation}\label{error1}
\triangle \omega=\frac{|\omega_{BSC}-\omega^{*}|}{\omega_{BSC}},
\end{equation}
and
\begin{equation}\label{error2}
\triangle L_x=\frac{|L_{x}-L_x^{*}|}{L_{x}},
\end{equation}
with $\omega^*$, and $L_x^{*}$ as the frequency and the length of the cavity corresponding to the degeneracy point in which
the BSC would occur according the two-mode approximation Eq. (\ref{two_mode}). For BSC $\it 3$ we obtained $\triangle \omega=0.019$, and $\triangle L_x=0.018$.
In Fig. \ref{BSC1} we also plot the BSC shape function constructed by Eq. (\ref{Shape_function}). One can see from Fig. \ref{BSC1}
that the two-mode approximation reproduces our numerical result to a good accuracy.
We also found that the same holds true for the other four BSCs in Fig. \ref{Trans}.
Therefore, the other BSC functions are not presented in this paper.

\begin{figure}
\begin{center}
\includegraphics[width=0.8\textwidth]{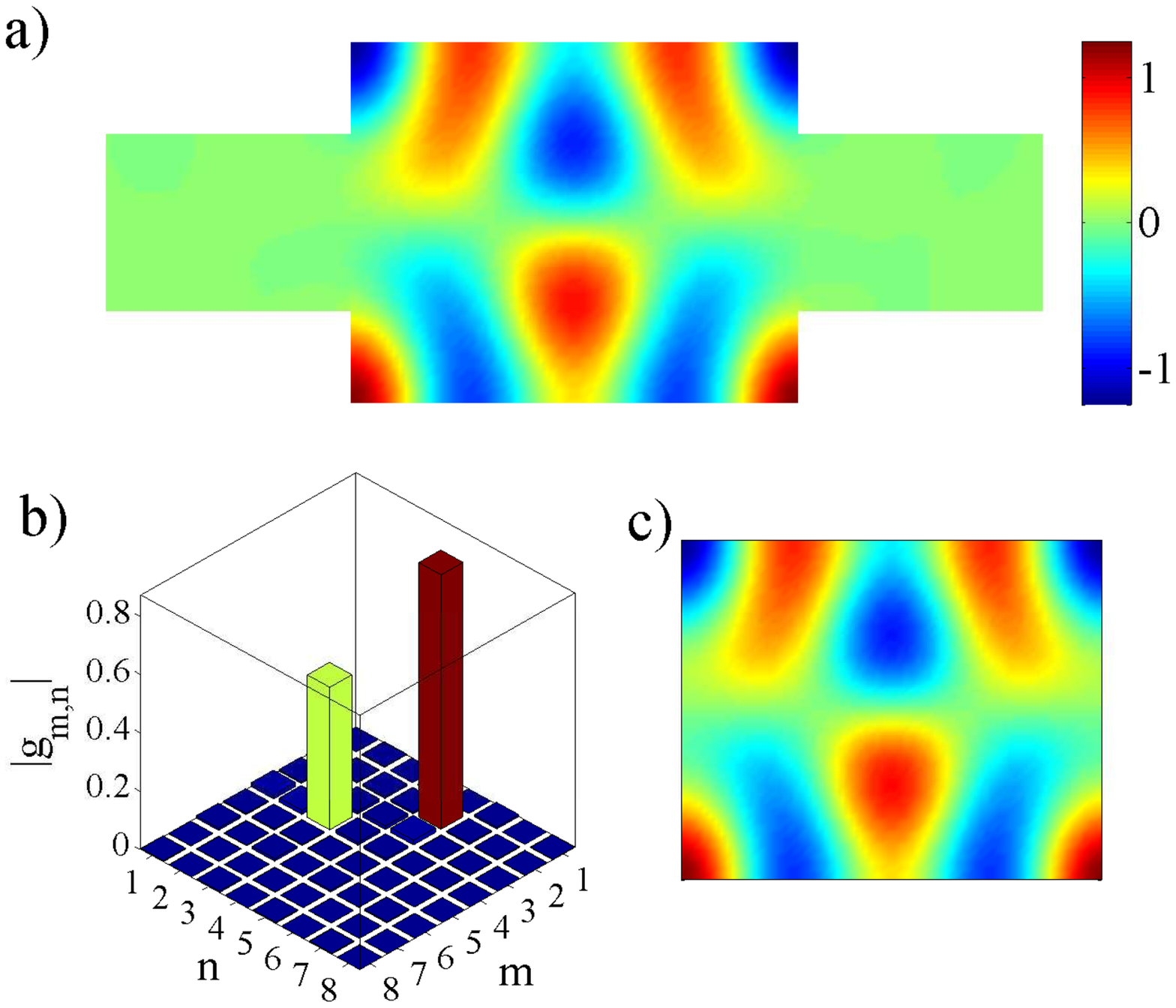}
\caption{(Color on-line) Bound state in the continuum antisymmetric about the duct axis of a two-dimensional duct-cavity structure
$\omega_{BSC}=5.385,  L_y=2, L_x=2.441$; a) Pressure field of the
 numerically computed BSC shape function; b) Absolute values of the modal coefficient of the numerically computed pressure field;
 c) BSC shape function found from the two-mode approximation Eq. (\ref{Shape_function}).}
\label{BSC2}
\end{center}
\end{figure}

\begin{figure}
\begin{center}
\includegraphics[width=0.8\textwidth]{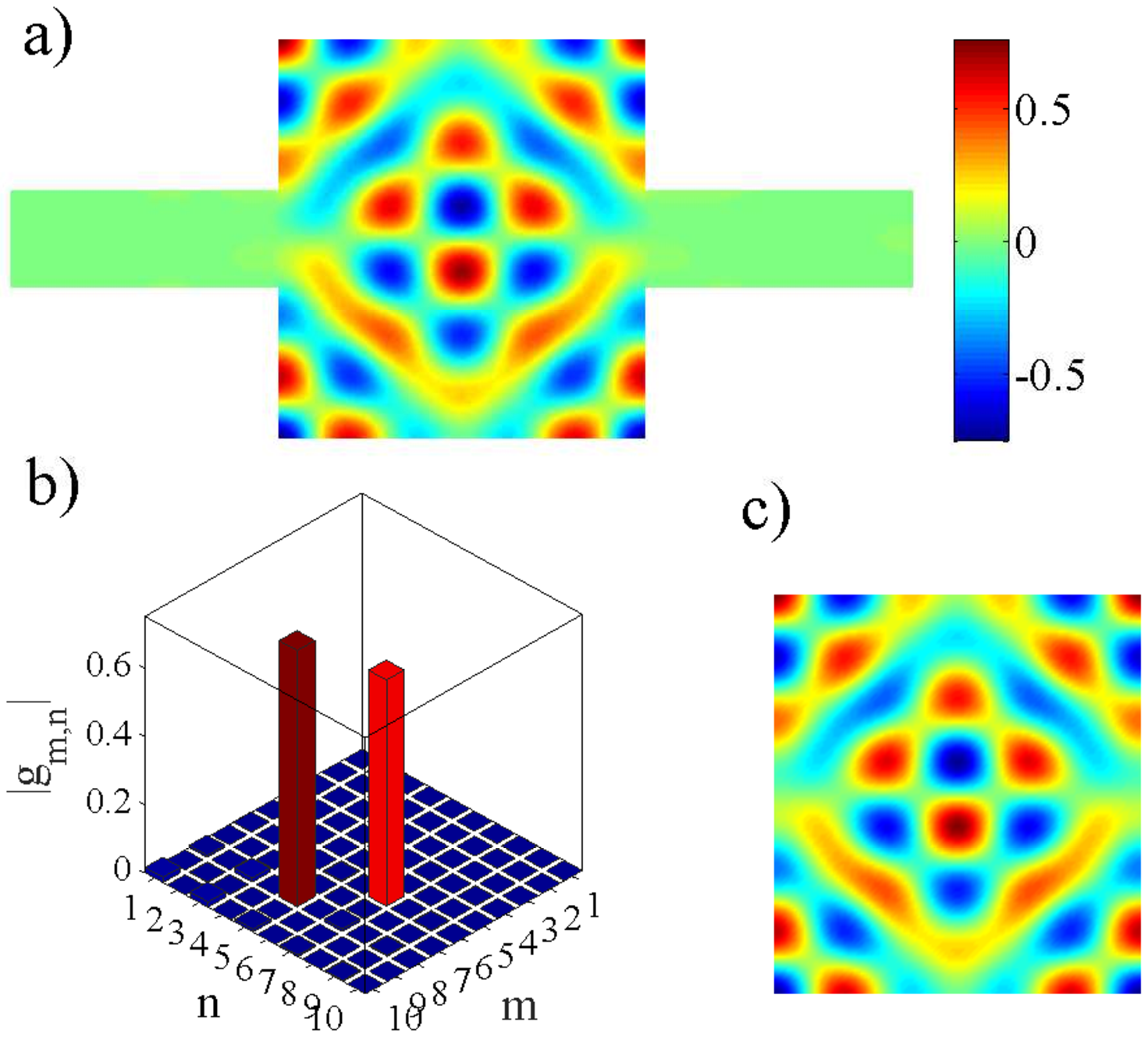}
\caption{(Color on-line) Bound state in the continuum antisymmetric about the duct axis of a two-dimensional duct-cavity structure
$\omega_{BSC}=6.502,  L_y=4, L_x=3.627$; a) Pressure field of the
 numerically computed BSC shape function; b) Absolute values of the modal coefficient of the numerically computed pressure field;
 c) BSC shape function found from the two-mode approximation Eq. (\ref{Shape_function}).}
\label{BSC_large}
\end{center}
\end{figure}

The same procedure was repeated in case of the incoming wave in the first antisymmetric channel for cavities with $L_y=2$.
As before we observed that the two-mode approximation
is valid in the situation when only one antisymmetric channel is open in each waveguide. For brevity we omit a
detailed description of our findings restricting ourselves to one example of a duct axis antisymmetric BSC shown
in Fig. \ref{BSC2}.
The $y$-symmetric BSC in Fig. \ref{BSC2} is formed by eigenmodes $\psi_1=\psi_{5,2}(x,y)$, and
$\psi_2=\psi_{3,4}(x,y)$ with the corresponding coupling constants $v_1=v^{5,2}_{L,1}=v^{5,2}_{R,1}=-0.692$, and
$v_2=v^{3,4}_{L,1}=v^{3,4}_{R,1}=0.384$. For the corresponding relative errors Eqs. (\ref{error1},\ref{error2}) we found
$\triangle \omega=0.004$, and $\triangle L_x=0.004$. As it was mentioned in Section 2, we expect that the two-mode approximation
could be applied irrelevant to the cavity size. This is confirmed by numerical tests for larger cavities $L_y=4$. As an example in Fig.
\ref{BSC_large} we present the duct-axis antisymmetric BSC formed by eigenmodes $\psi_1=\psi_{5,8}(x,y)$, and
$\psi_2=\psi_{7,6}(x,y)$ with the coupling constants $v_1=v^{5,8}_{L,1}=v^{5,8}_{R,1}=-0.371$, and
$v_2=v^{7,6}_{L,1}=v^{7,6}_{R,1}=0.402$. The values of the relative errors $\triangle \omega=0.003$, and $\triangle L_x=0.007$
indicate that the two-mode approximation holds true for larger cavities.

\begin{figure}
\begin{center}
\includegraphics[width=0.8\textwidth]{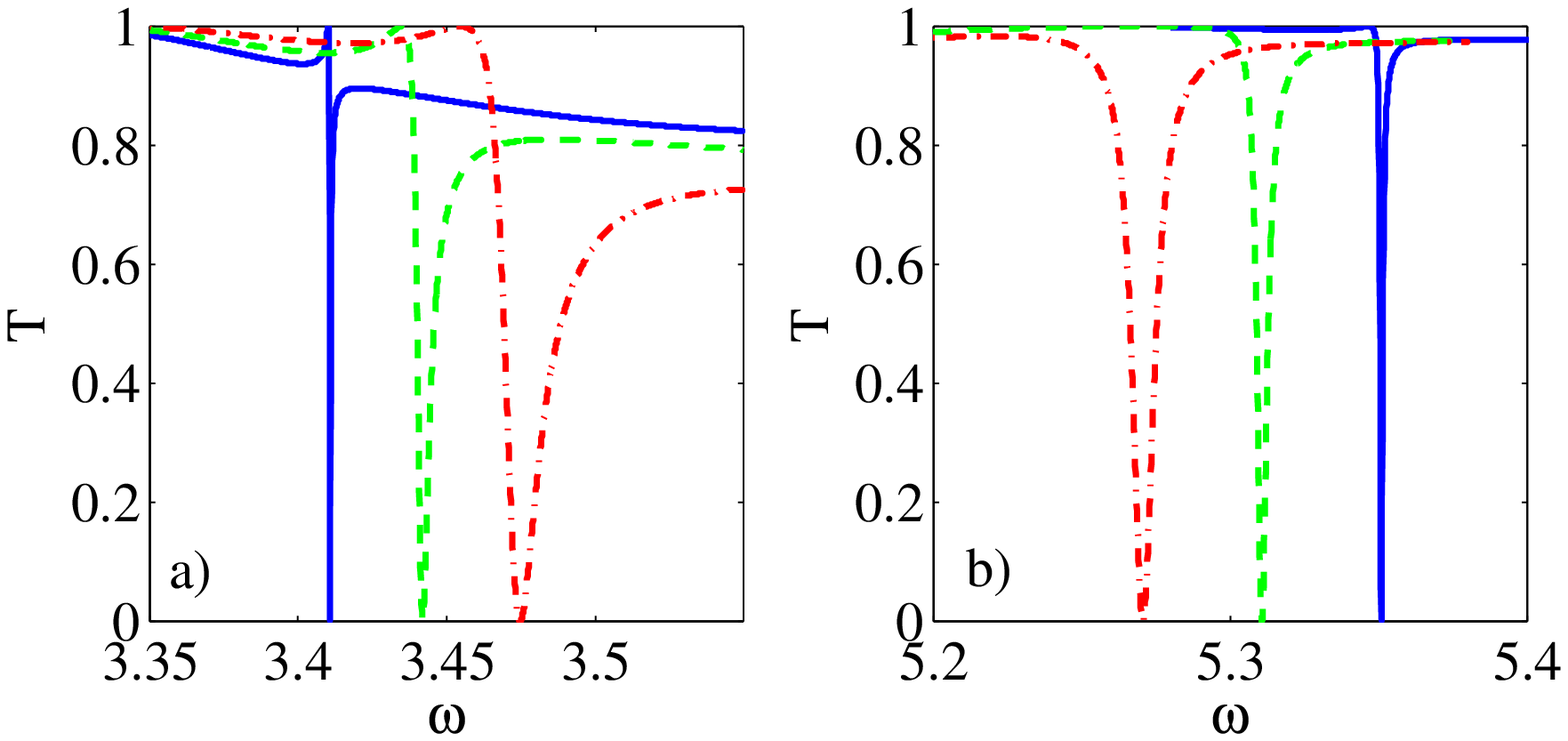}
\caption{(Color on-line) Collapse of Fano resonance in the vicinity of a BSC point under variation of the length of
the resonator $L_x$. a) BSC {\it 4} in Fig. \ref{BSC1} $\omega_{BSC}=3.396, L_y=2, L_x=2.776$. Resonator lengths:
 $L_x=2.75$ - solid blue line,
 $L_x=2.70$ - dashed green line, $L_x=2.60$ - dash-dot red line.
 b) Duct axis antisymmetric BSC in Fig. \ref{BSC2} $\omega_{BSC}=3.384, L_y=2, L_x=2.441$. Resonator lengths: $L_x=2.460$ - solid blue line,
 $L_x=2.485$ - dashed green line, $L_x=2.510$ - dash-dot red line.}
\label{Fano_2D}
\end{center}
\end{figure}

\begin{figure}
\begin{center}
\includegraphics[width=0.8\textwidth]{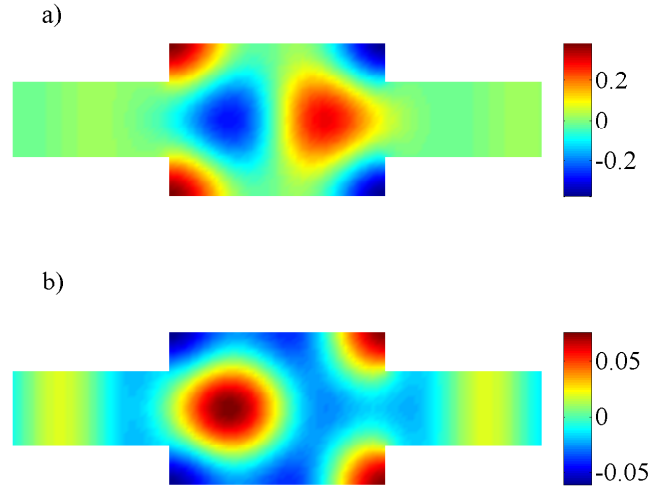}
\caption{(Color on-line) Scattering function at the peak $\omega=3.410$ of the Fano resonance Fig. \ref{Fano_2D} related to BSC {\it 3}
shown in Fig. \ref{BSC1}.
$L_y=2, L_x=2.75$. a) Real part of the scattering function. b) Imaginary part of the scattering function.}
\label{Scattering_function}
\end{center}
\end{figure}

As it was mentioned in the introduction the occurrence of BSC is associated with
a collapse of Fano resonance. To illustrate this in Fig. \ref{Fano_2D} we plot the dependance transmittance vs.
frequency for three different lengths of the resonator $L_x$ close to the BSCs point. It was recently demonstrated by
\cite{Bulgakov2014} that if the BSC point is approached in the parametric space along the certain path corresponding
to the peak of the Fano resonance, the long-lived complex mode related to the BSC provides the dominant contribution to the
scattering function. In consequence of this the scattering function near the BSC point bears the maximal resemblance
of the BSC if the parameters are tuned to the peak of the Fano resonance. This is illustrated in Fig. \ref{Scattering_function}
where one can see that the real part of the scattering function is almost identical to the BSC from Fig. \ref{BSC1} meanwhile
the imaginary is present as a background providing the net power flux in the system. Thus, one can experimentally observe
a signature of BSC in the form of a large amplitude acoustic resonance even though in a realistic set-up the parameters
are always detuned from the true BSC point due, for example, to fabrication inaccuracies.

\section{Bound states in the continuum in three-dimensional open cavities}

\begin{figure}
\begin{center}
\includegraphics[width=0.8\textwidth]{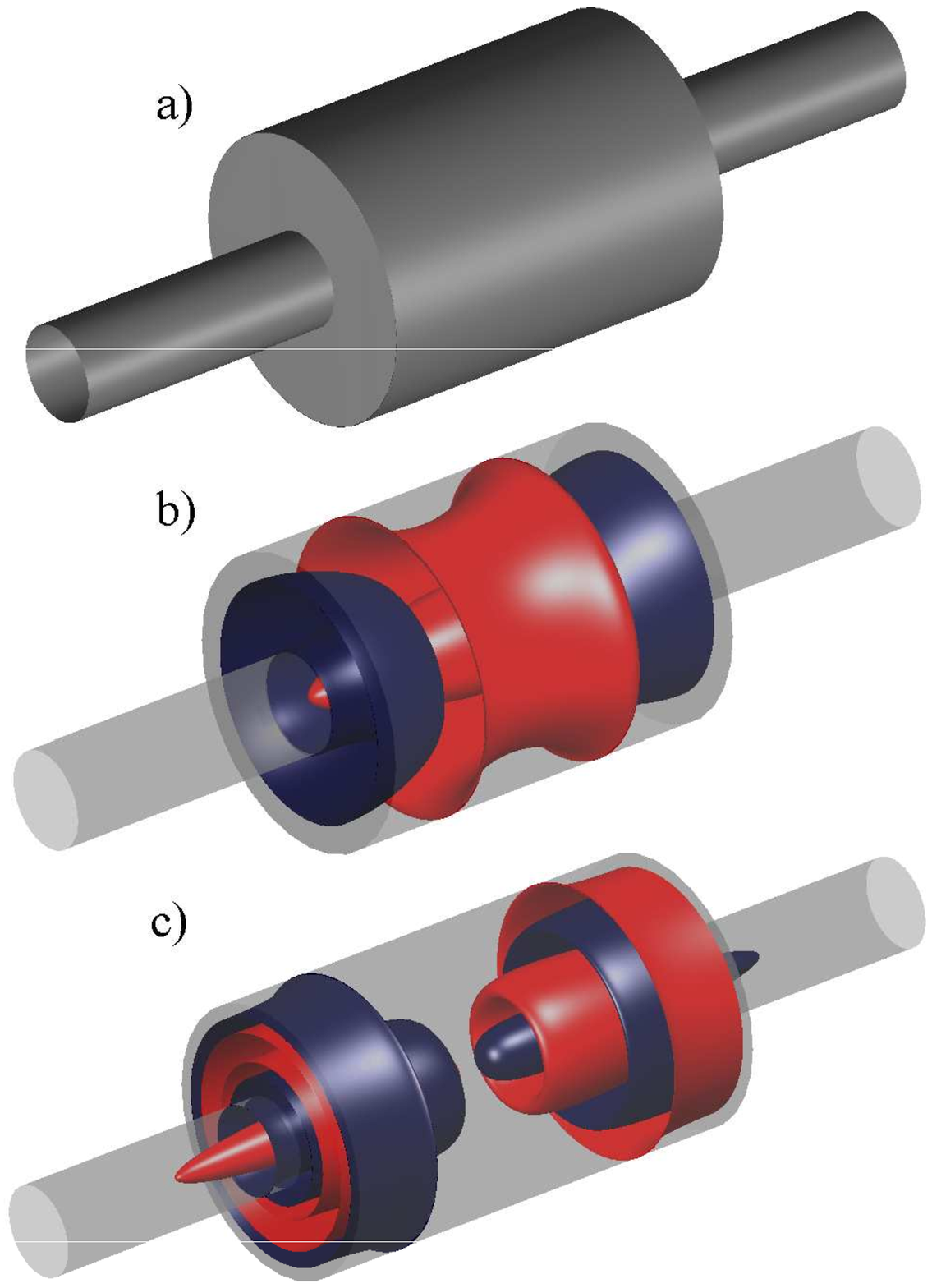}
\caption{Bound states in axisymmetric duct-cavity structure. a) Cylindrical resonator of radius $R$ and length $L_x$ with two coaxially attached waveguides
of radius $a=1$. b)
Pressure field isosurfaces of BSC $\it 2$ in Fig. \ref{BSC3} symmetric with respect to the central section of the resonator;
 dark blue
$\psi_{BSC}=-0.1$, light red $\psi_{BSC}=0.1$. c) Pressure field isosurfaces of BSC $\it 7$ in Fig. \ref{BSC4} antisymmetric with respect
 to the central section of the resonator; dark blue
$\psi_{BSC}=-0.1$, light red $\psi_{BSC}=0.1$.}
\label{3dpic}
\end{center}
\end{figure}
\begin{figure}
\begin{center}
\includegraphics[width=0.8\textwidth]{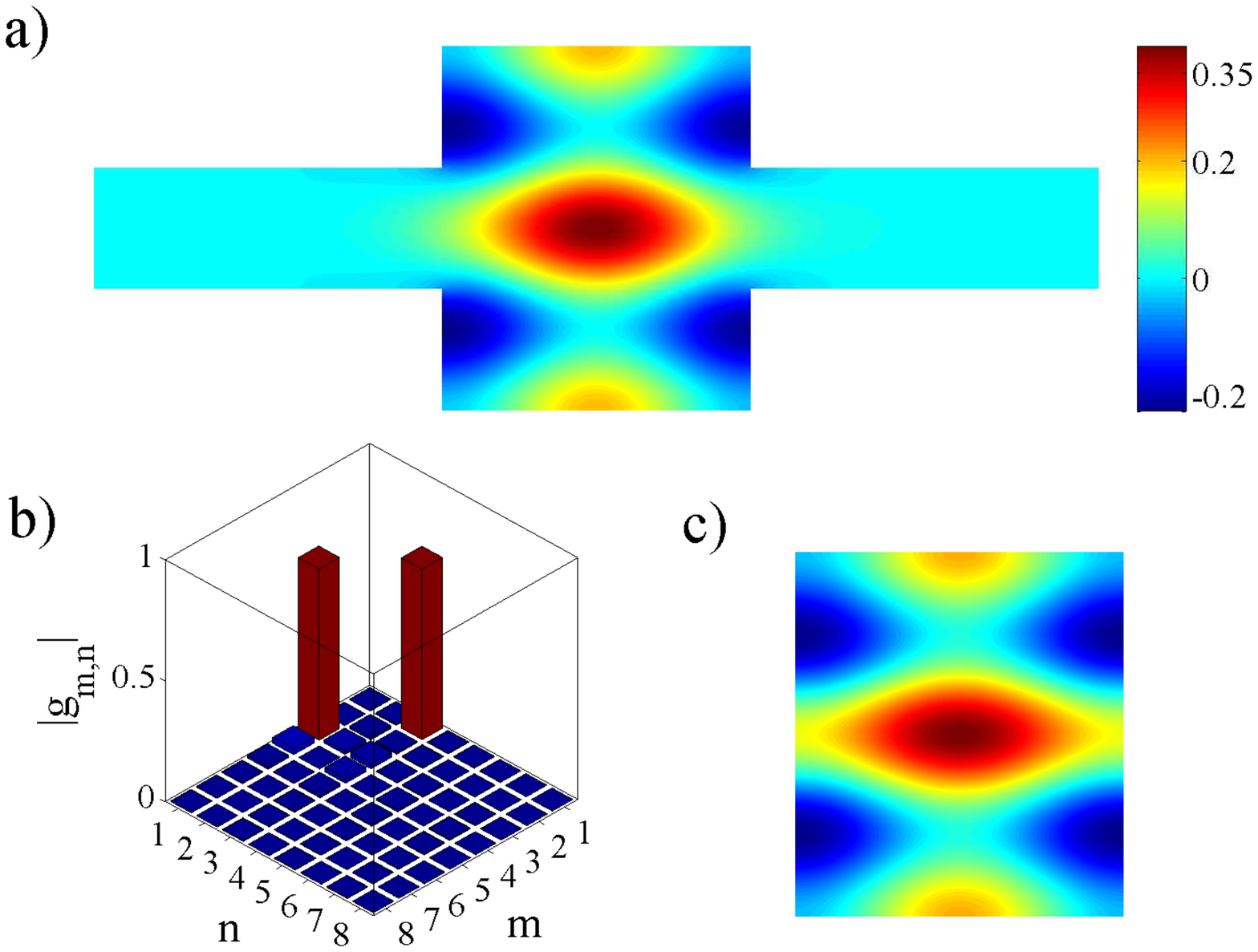}
\caption{(Color on-line) BSC $\it 2$ in a three-dimensional duct-cavity structure
$\omega_{BSC}=2.367, R=2, L_x=2.655$; a) $x0y$-plane section of the pressure field of the
 numerically computed BSC shape fnction; b) Absolute values of the modal coefficient of the numerically computed pressure field; c)
 BSC shape function found from the two-mode approximation Eq. (\ref{Shape_function}).}
\label{BSC3}
\end{center}
\end{figure}
\begin{figure}
\begin{center}
\includegraphics[width=0.8\textwidth]{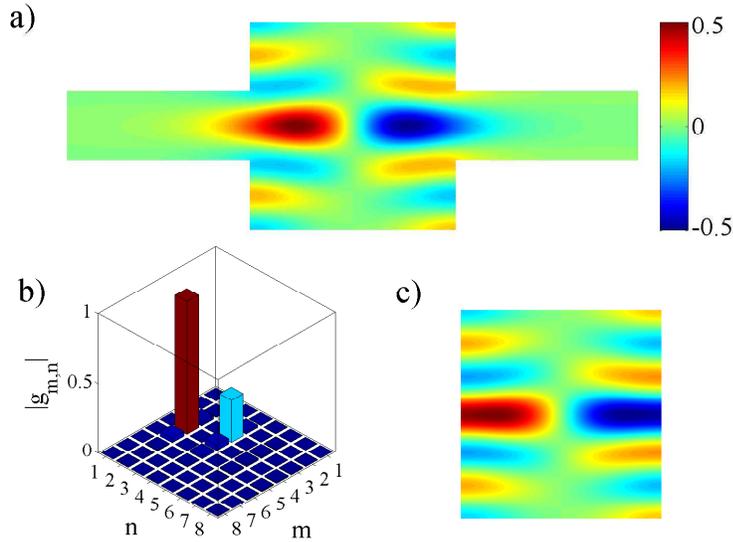}
\caption{(Color on-line) BSC $\it 7$ in a three-dimensional duct-cavity structure
$\omega_{BSC}=3.557, R=2, L_x=3.387$; a) $x0y$-plane section of the pressure field of the
 numerically BSC; b) Absolute values of the modal coefficient of the numerically computed pressure field;
  c) BSC shape function found from the two-mode approximation Eq. (\ref{Shape_function}).}
\label{BSC4}
\end{center}
\end{figure}

Let us consider the axisymmetric
cylindrical resonator with two coaxially attached cylindrical waveguides of radius $a=1$
as depicted in Fig. \ref{3dpic},a). After the azimuthal variable is separated we can restrict ourself to
the case of zero angular momentum bearing in mind that our results can be easily generalized to the non-zero case
taking into account that by virtue of Eq. (\ref{coupling}) only the modes of the same rotational number are coupled on
the waveguide-cavity interface.
\begin{figure}
\begin{center}
\includegraphics[width=1\textwidth]{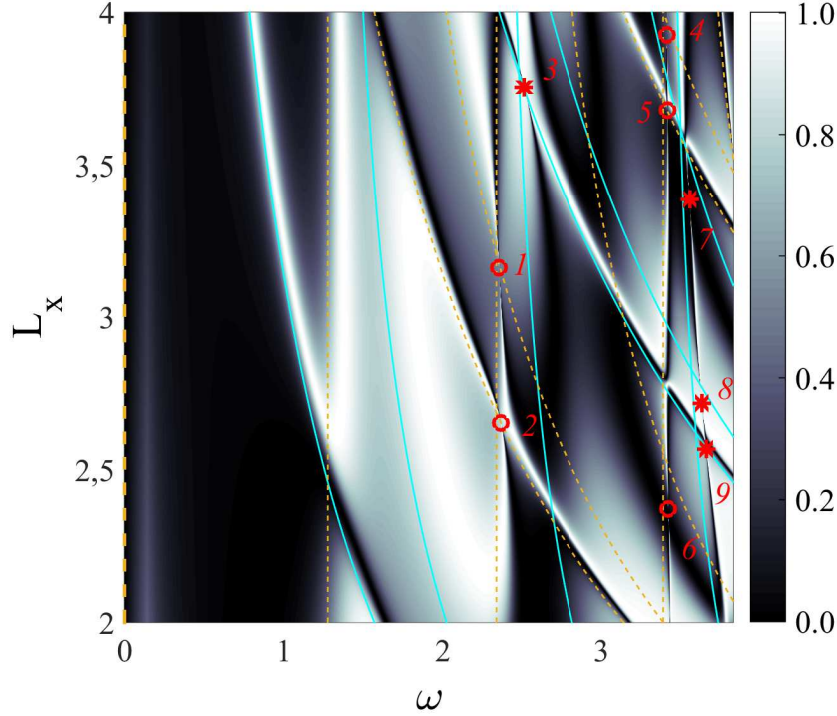}
\caption{(Color on-line) Transmittance for a cylindrical resonator with the radius $R=2$ with two coaxially attached cylindrical
waveguides of radius $a=1$ vs. frequency $\omega$ and the length of the resonator $L_x$. The dashed lines show eigenfrequencies of
the modes symmetric about
the $y0z$-plane; solid line - antisymmetric ones. The positions of the BSCs are shown by open circles and stars for $y0z$-symmetric and $y0z$-antisymmetric
modes, correspondingly.}
\label{Trans3D}
\end{center}
\end{figure}
In our case the cavity eigenfrequencies are given by
\begin{equation}\label{eig3D}
\omega_{m,n}^2=\left[\frac{\mu_{m}^2}{R^2}+\frac{\pi^2 (n-1)^2}{L_x^2}\right]
\end{equation}
where $R$ and $L_x$ are the radius and length of the cylindrical resonator, and $\mu_m$ is the m-th root of the
equation $J_0'(\mu_m)=J_1(\mu_m)=0, m=1, 2, \ldots$ for the derivative of the first order Bessel function. The corresponding eigenfunctions are
\begin{equation}\label{eigfun3D}
    \psi_{m,n}(r,x)=\frac{1}{\sqrt{\pi}RJ_0(\mu_m)}J_0\left(\frac{\mu_mr}{R}\right)\psi_n(x),
\end{equation}
where $r$ is the radial coordinate in the $y0z$-plane while the axial modes are given by
\begin{equation}
\psi_n(x)=\sqrt{\frac{2-\delta_n^{1}}{L_x}}\cos\left(\frac{\pi (n-1) (2x+L_x)}{2L_x}\right), \ n=1, 2, \ldots
\end{equation}
As before the transmission problem was solved by the use of the acoustic coupled mode theory \cite{Maksimov2015}
and the complex eigenvalues were obtained by the harmonic inversion method. The frequency $\omega$ varies in the range $[0, \mu_2], \mu_2=3.831$
which means that only one propagating channel is open in each waveguide.
The transmittance along with the spectrum of the closed resonator and the positions of BSCs is shown in Fig. \ref{Trans3D}. The complex eigenfrequencies
are plotted in Fig. \ref{Complex_eigenfrequencies_3D} against the length of the resonator $L_x$.  In Figs. \ref{BSC3},\ref{BSC4}
we demonstrate the shape functions of two BSCs occurring through the two mode interference mechanism.
BSC {\it 2} in Fig. \ref{BSC3} with $\triangle \omega=0.012$, and $\triangle L_x=0.012$ is formed by eigenmodes
$\psi_1=\psi_{1,3}(r,x)$, and
$\psi_2=\psi_{3,1}(r,x)$ with the corresponding coupling constants $v_1=v^{3,1}_{L,1}=v^{3,1}_{R,1}=0.310$, and
$v_2=v^{1,3}_{L,1}=v^{1,3}_{R,1}=0.289$, while $y0z$-antisymmetric BSC {\it 7} depicted in Fig. \ref{BSC4} is formed
by $\psi_1=\psi_{2,4}(r,x)$, and
$\psi_2=\psi_{4,3}(r,x)$ with the coupling constants $v_1=v^{2,4}_{L,1}=-v^{2,4}_{R,1}=-0.078$, and
$v_2=v^{4,3}_{L,1}=-v^{4,3}_{R,1}=0.389$.  A three dimensional visualization of BSCs from Figs. \ref{BSC3},\ref{BSC4} is offered
in Fig. \ref{3dpic} b,c) in form of pressure field isosurfaces. One can see that of all BSCs in Fig. \ref{Trans3D} BSC {\it 7} deviates
the most from the corresponding degeneracy point $\triangle \omega=0.016$, and $\triangle L_x=0.066$. The reason for this can be understood from Figs. {\ref{3dpic}} c), \ref{BSC4}
where one can clearly see that
BSC {\it 7} extends far into the waveguides due to the coupling to the evanescent modes which is neglected by the two mode approximation.
One can conclude that the coupling to the evanescent modes becomes important just below the next channel propagation threshold
as the channel function extends far into the waveguide.
Nevertheless, the two-mode approximation is still valid in the three-dimensional case reproducing the numerically computed shape functions
to a good accuracy.
\begin{figure}
\begin{center}
\includegraphics[width=0.8\textwidth]{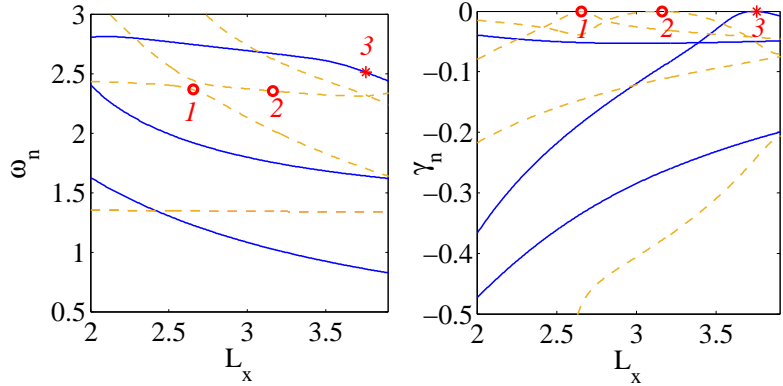}
\caption{(Color on-line)  Complex eigenfrequencies against the length of the three-dimensional resonator $L_x$; real parts - left panel,
imaginary parts - right panel. The $y0z$-symmetric resonances are shown by dash brown lines; $y0z$-antisymmetric by blue
solid lines.  The positions of BSCs from Fig. \ref{Trans3D} are shown by red open circles and stars for $y0z$-symmetric and
$y0z$-antisymmetric modes, correspondingly.}
\label{Complex_eigenfrequencies_3D}
\end{center}
\end{figure}

In this paper, we restricted ourself to coaxial systems in which the azimuthal number is a constant of motion
so that the rotational vibration modes do not participate in formation of BSC depicted in Figs. \ref{BSC3}, \ref{BSC4}.
It is worth noticing, however, that the cavity eigenmodes with azimuthal number $\nu\neq 0$ are symmetry
protected BSCs as their eigenfrequencies are below the $\nu$-th propagating threshold in cylindrical waveguides.

\section{Conclusion}

We considered sound transmission through two- and three-dimensional acoustic resonators
and demonstrated the existence of singular points in which
the unit transmittance coalesces with the zero transmittance in the points of collapse of Fano resonances.
These singular points occur due to a full destructive interference of two degenerate  modes
of the same symmetry. As a result a certain superposition of these two modes is a
trapped solution corresponding to a bound state in the continuum or embedded trapped mode localized within the resonator.
The above mechanism of
wave localization was first proposed by \cite{Friedrich1985}, and has been so far experimentally realized only in
microwave set-ups \cite{Lepetit1, Lepetit2014}.
In the present paper we invited the reader's attention to acoustic cylindrical resonators coaxially connected to
semi-infinite waveguides where the degeneracy points could be accessed in the parametric space under variation of the length of the resonator.
We speculate that such a tuning could be performed in a realistic acoustic experiment
by the use of piston-like hollow-stem waveguides tightly fit to the interior boundaries of a cylindric cavity.
For analysing the bound states in the continuum we successfully employed the acoustic coupled mode theory \cite{Maksimov2015}
which allowed us to analytically predict the eigenfrequencies of the bound states as well as their shape functions.
Until recently \cite{Hein2012} the bound states in the continuum were independently studied in physics and fluid mechanics with
little to none cross-fertilization. In the present paper we attempted to bridge that gap to the mutual benefit of the both
communities.

{\bf Acknowledgement}: This work has been supported by Russian Scientific Foundation through Grant 14-12-00266.

\bibliographystyle{jfm}
\bibliography{bound_states}
\end{document}